
\input phyzzx
\doublespace
 \newtoks\slashfraction
 \slashfraction={.13}
\def\diag#1#2#3{\left( \matrix{#1&\ &0\cr
                               \ &#2&\ \cr
                               0&\ &#3\cr} \right) }
\def\vector#1#2#3{\left( \matrix{#1\cr #2\cr #3\cr} \right) }

\def\uem{{\rm U(1)_{em}}}
\def\zt{{\rm{\bf Z}_2}}
\def\ztwo{{\rm{\bf Z}_2}}
\def\uz{{\uem \times \ztwo}}
\def\pint{{\rm Pin(2)}}
\def\uone{{\rm U(1)}}
\def\sut{{\rm SU(2)}}
\def\suthree{{\rm SU(3)}}
\def\ot{{\rm O(2)}}
\def\sothree{{\rm SO(3)}}
\def\dt{{\rm D_2}}
\def\zf{{\rm{\bf Z}_4}}
\def\q{{\rm Q}}
\def\G{{\rm G}}
\def\H{{\rm H}}

\pubnum={SLAC-PUB-5805}
\pubtype={T}
 \date{April 1992}
\titlepage
\title{Superconductivity Solves the Monopole Problem for Alice Strings
\foot{Supported by the Department of Energy, contracts
 DE-AC03-76SF00515 and DE-FG03-84ER-40168.}}
\author{Shahar Ben-Menahem}
\SLAC
\author{Adrian R.~Cooper}
\address{Department of Physics\break
University of Southern California\break
Los Angeles, California 90089-0484}
\medskip
\abstract
Alice strings are cosmic strings that turn matter into antimatter.
Although they arise naturally in many GUT's, it has long been believed
that because of the monopole problem they can have no cosmological effects.
We show this conclusion to be false; by using the Langacker-Pi mechanism,
monopoles can in fact be annihilated while Alice strings are left intact.
This opens up the possibility that they can after all contribute
to cosmology, and we mention some particularly important examples.

\submit{Nuclear Physics \bf B}
\endpage
\chapter{Introduction.}

\REFS\schwarz{A.S.~Schwarz \journal Nucl.Phys.&B208 (82) 141; \hfil \break
A.S.~Schwarz and Y.S.~Tyupin \journal Nucl.Phys.&B209 (82) 427.\hfil }
\REFSCON\preskill{J.~Preskill and L.~Krauss \journal Nucl.Phys.&B341
(90) 50; \hfil \break
M.~Bucher, H-K.~Lo, and J.~Preskill, ``Topological Approach
to Alice Electrodynamics,'' Caltech Preprint CALT-68-1752 (1992).\hfil }
\REFSCON\coleman{M.~Alford, K.~Benson, S.~Coleman, J.~March-Russell,
and F.~Wilczek \journal Nucl.Phys. \hfil \break
&B349 (91) 414.\hfil }
\REFSCON\brekke{L.~Brekke, W.~Fischler, and T.~Imbo, ``Alice Strings, Magnetic
Monopoles, and Charge Quantisation,'' Harvard Preprint HUTP-91/A042 (1991).}
\REFSCON\langpi{P.~Langacker and S.-Y.~Pi \journal Phys.Rev.Lett. &45 (80)
1.}
\REFSCON\tubebreak{A.E.~Everett, T.~Vachaspati, and A.~Vilenkin
\journal Phys.Rev.D &31 (85) 1925; \break
E.~Copeland, D.~Haws, T.W.B.~Kibble, D.~Mitchell, and N.~Turok
\journal Nucl.Phys. \hfil \break
&B298 (88) 445. \hfil }
\REFSCON\kibbwein{T.W.B.~Kibble and E.J.~Weinberg \journal Phys.Rev.D &43
(91) 3188.}
\REFSCON\preswise{J.~Preskill, S.P.~Trivedi, F.~Wilczek, and M.B.~Wise
\journal Nucl.Phys \hfil \break
&B363 (91) 207.}
\REFSCON\farris{T.H.~Farris, T.W.~Kephart, T.J.~Weiler, and T.C.~Yuan,
``The Minimal Electroweak Model for Monopole Annihilation,'' Vanderbilt
Preprint VAND-TH-91-06 (October 1991).}
\REFSCON\efficient{
R.~Holman, T.W.B.~Kibble, and S.-J.~Rey, ``How Efficient is the
Langacker-Pi Mechanism of Monopole Annihilation?'' Santa-Barbara Preprint
NSF-ITP-09-92 (1992).}
\REFSCON\numsim{
M.~Hindmarsh and T.W.B.~Kibble \journal Phys.Rev.Lett &55 (85) 2398;
\hfil \break
T.~Vachaspati and A.~Vilenkin \journal Phys.Rev.D &35 (87) 1131; \hfil \break
S.H.~Lee and E.J.~Weinberg \journal Phys.Rev.D &42 (90) 3422; \hfil \break
B.~Allen and R.R.~Caldwell \journal Phys.Rev.D &43 (91) 3173.}
\REFSCON\dogpiss{
S.~Ben-Menahem and A.R.~Cooper, in preparation.}
\refsend

Alice strings [\schwarz,\preskill,\coleman] are a class of cosmic string
with the remarkable property that a particle travelling around one will
come back as its own antiparticle.  They may be formed in any
Grand Unified Theory in which
the charge conjugation operator is contained in the original gauge group.
This occurs, for example, in the standard ${\rm SO(10)}$ GUT.
The existence of such objects could obviously
have dramatic consequences for cosmology.  It has long been appreciated
[\brekke] that a simple spontaneous symmetry breaking
pattern that gives
rise to Alice strings will also create magnetic monopoles.  Since the monopole
density of the present Universe is known to be very low, some mechanism must
exist for getting rid of them.  The most common solution is to claim that the
pertinent phase transition occured before some inflationary era.
The monopoles would then have been swept outside our horizon, easily satisfying
any observed bounds on their density.  However, a feature of
this solution is that
inflation would have erased any other topological defects in exactly the same
way.  In particular, since Alice strings were produced at the same phase
transition, none would survive to cosmologically interesting times.

There is, however, an alternative explanation for the scarcity of present day
monopoles due to Langacker and Pi [\langpi].  In this, the Universe enters
a temporary superconducting phase after the original phase transition.
This causes the magnetic flux to be confined to tubes that end at
(anti) monopoles.  The high tension in these tubes causes monopoles to
annihilate extremely quickly.  This rate may be increased even further
by the tendency of the flux tubes to break, forming monopole-antimonopole pairs
along their length [\tubebreak,\kibbwein].
Annihilation then need only occur between
neighbouring particles.
When the superconducting phase is exited, the monopole density has been
reduced to acceptable values, which may be far less than one per
horizon volume [\kibbwein].

At first glance, it may seem that the Langacker-Pi mechanim is rather
contrived; we might worry about how elaborate a theory must be in order
to give an intermediate superconducting phase.  Surprisingly, though,
it turns out that even a simple extension of the Standard Model
with one additional charged scalar is
capable of producing this phase structure [\farris].  The Langacker-Pi
mechanism must therefore be taken very seriously as a potential solution
to the monopole problem.

In the present paper we examine the effects of this mechanism in a model
containing Alice strings.  We find that breaking $\uz$
\foot{Here $\zt$ is the discrete group $\{ 1,{\rm C} \}$ generated
by the charge conjugation operator C.}
to different discrete subgroups leads to different networks of flux tubes,
which contain both Alice strings and magnetic flux lines.
For our purposes, there are three categories into which such networks fall,
and we analyze them using simple toy models.  In the first case, $\uz$ is
partially broken such that each magnetic monopole has at least two flux
tubes attached to it.  Because these tubes will generally be pulling the
monopoles in different directions, the annihilation efficiency is fairly
low [\numsim], and it is not known whether it can occur fast enough to
solve the monopole problem.
In the second case, $\uz$ is completely broken, which means that the
flux tubes all have trivial holonomy.  If the hierarchy
\foot{The term ``hierarchy'' refers to the energy-scale difference between
the monopole-forming and the flux-tube-forming phase transitions.
That is, it defines the relative scales of the monopole mass and the flux
tube tension.}
is large, then as before a network of strings will form.  Although
each monopole is attached to just one flux tube, there will also be
loops of string that are multiply magnetically charged.  These will not be
neutralised quickly, and so the usual Langacker-Pi mechanism will be
evaded.
The efficacy of monopole annihilation in either of these cases is not
yet known and a verdict must await more thorough studies of network
evolution.

For the third case, we introduce a new model containing Alice strings.
In this, it is possible to break $\uem$ completely while leaving $\ztwo$
unbroken.  If this is done at a low hierarchy, then we can show that
monopoles annihilate much faster than the
rest of the network.  Thus the monopole bounds may be satisfied while leaving
behind a high density of Alice strings.  A variant of this
mechanism, motivated by the work of Kibble and Weinberg [\kibbwein], enables
us to prevent monopoles from forming in the first place.

We conclude that it is possible for the pertinent phase transition to occur
{\it after} any inflationary era.  This means that Alice strings may
contribute interesting effects to cosmology, particularly after the
superconducting period.  Of course we know that today C is {\it not}
a symmetry of the vacuum.  This is not a problem, since a necessary
condition for a model to contain Alice strings is that C be a member of the
original gauge group.  It can later be spontaneously broken (together with CP),
whereupon Alice strings become the
boundaries of, possibly superconducting, domain walls.  String-bounded walls
eventually decay, and can have interesting consequences without dominating
the energy density of the Universe [\preswise].
In a future publication we discuss the cosmological implications
of Alice strings
further, and show in particular how these domain walls could account for
the baryon asymmetry [\dogpiss].

\chapter{Alice Strings and the Monopole Problem}

In this section we introduce the simplest model containing Alice strings and
show how it also leads to monopoles.
This consists of a
(3+1)-dimensional non-abelian
theory with gauge group $\G=\sothree$
and a Higgs field $\Phi$ in the 5-dimensional
irreducible representation.  We can regard $\Phi$ as a real symmetric traceless
$3 \times 3$ matrix transforming as
$$\Phi \rightarrow M \Phi M^{-1} \qquad {\rm for}\ M \in \sothree \,.
\eqn\phitrans$$
The Higgs potential is chosen so that $\vev {\Phi}$ acquires two degenerate
eigenvalues.  In the unitary gauge,
$$\vev {\Phi}=p \diag 1 1 {-2}
\,, \eqn\phivev$$
where $p$ is some constant.
The unbroken subgroup of $\sothree$ is then
$\H=\uone \times_{\rm SD} {\bf Z}_2 \equiv \ot$,
and this is disconnected.
\foot{The subscript SD denotes a semi-direct product.}
Since we wish to interpret the $\uone$
factor as being
electromagnetism, it is clear that $\H$ also contains
the charge-conjugation operator C.

As our interest lies in the topological structures of the theory, it
is more convenient to work
with a simply connected unbroken gauge group.  We can do this by
considering the double cover of the above model.  This has a gauge group
$\G=\sut$, which is broken by a Higgs field $\Phi$ in the spin-$2$
representation.  As before, $\Phi$ may be regarded as a real traceless
symmetric $3\times 3$ matrix that transforms as in equation \phitrans\ and
acquires the expectation value \phivev .
The unbroken subgroup is now $\H=\pint$, where $\pint$ is the double cover
of $\ot$ and may be parametrised as
$$\pint=\{ e^{i {\theta \over 2} \sigma_3} ,
i \sigma_2 e^{i {\theta \over 2} \sigma_3} \} ,
\quad \theta \in [0,4\pi) \,. \eqn\pintdef $$
Because $\G=\sut$ is simply-connected, we have $\Pi_1(\G / \H) = \Pi_0(\H)$,
and also $\Pi_2(\G / \H)=\Pi_1(\H)$.

Since $\Pi_1 (\G / \H) \ne 0$, the theory admits topologically stable
cosmic strings.  The result of parallel transport around such a string defines
its ``magnetic flux'', $\Omega$, an element of $\H$:
$$\Omega=P {\rm exp} \left( i \ointop {\bf A} \cdot {\rm d} {\bf x} \right)
\in \H \,. \eqn\partrans$$
For a string whose flux lies in the
component of $\H$ disconnected from the identity,
$\Omega$ does not commute with the charge operator
$Q\equiv {1\over 2} \sigma_3$ that generates $\uone \subset \pint$.
Rather,
$$\Omega Q \Omega^{-1}=-Q \,. \eqn\chargeflip$$
Hence, a particle circumnavigating
the string will have the sign of its charge flipped when it returns.
This behaviour defines an Alice string.
As explained in [\preskill, \coleman], electric and magnetic field lines
in the presence of an Alice string have their directions reversed as they
cross some gauge-dependent branch cut.

Notice that since $\Pi_2(\G / \H)=\Pi_1(\H) \ne 0$, the spontaneous symmetry
breaking also generates magnetic monopoles.  This behaviour is
generic [\brekke], and
in particular persists even
if $\G$ contains an explicit factor of $\uone$.
We may think that the situation can be saved in the usual way by simply
asserting that an era of inflation is entered after the $\G \rightarrow \H$
phase transition.  However, as the monopoles are swept outside the
horizon, so too will be any other topological structures such as Alice strings.
It is the purpose of this paper to provide a toy model in which the density
of monopoles is reduced to satisfy experimental bounds, while the
density of Alice strings remains high enough to contribute interesting
effects to cosmology.

\hfill

\chapter{Alice Strings in a Superconducting Universe}

In order to erase monopoles,
we shall require the Universe to enter a temporary superconducting
phase, following Langacker and Pi [\langpi].
To understand the effects of this, it is very instructive to
first consider the fate of monopoles and Alice
strings when the group $\pint$ is broken down to
some discrete subgroup.
In particular $\ot$
contains a discrete subgroup $\dt$.
This is the dihedral group, and consists of rotations of $\pi$ about the
principal axes, so
$$\dt=\{ 1,a,b,c \} \qquad {\rm where} \qquad a^2=b^2=c^2=1,
\quad ab=c, \quad ac=b, \quad bc=a \,.
\eqn\dtwodef$$
Now, $\ot \subset \sothree$, and as $\sothree$ is lifted to its double cover
$\sut$, then $\ot$ is lifted to $\pint$.  Similarly, $\dt$ is lifted
to the quaternionic group $\q$, where
$$\q=\{\pm 1,\pm i,\pm j,\pm k \}
\equiv \{\pm 1, \pm i \sigma_1, \pm i \sigma_2, \pm i \sigma_3 \}
\,. \eqn\qdef$$
If we start with the group $\sut$ and break it down to $\pint$, then,
as discussed in the previous section, we will generate both Alice strings
and magnetic monopoles.  Now we break the symmetry further to $\q$, and
let $k$ correspond to the generator of $\uem$.

As this breaking occurs, the flux emanating from a monopole will form
four tubes,
\foot{In the sequel we shall often use the terms ``string'' and
``flux tube'' interchangeably.}
 each with holonomy $k$, as shown in
\fig{The flux tubes associated with a magnetic monopole after
$\pint$ has been broken down to $\q$.}.
This is because the holonomy of any closed loop must now be an element of
$\q$.  Furthermore, it is energetically favourable for a flux tube to break
into the smallest allowed flux fractions.
Since $\pm i \sigma_1$ and $\pm i \sigma_2$ correspond to charge conjugation
operators in the group $\pint$, it should be clear that tubes carrying
quaternionic flux $i$ or $j$ are the remains of Alice strings.

In addition to monopoles at which four $k$-tubes meet, there will be
vertices at which four $i$-tubes meet; we call them $i^4$ vertices.
These correspond to the joining
of four Alice strings.  An $i^2k^2$ vertex describes
a half-charged monopole
\foot{It is essential in any Grand Unified Theory that fractionally
charged monopoles cannot exist by themselves.  The objects that we are
considering here, though, can never be separated from the Alice string,
and so they are allowed.}
threaded onto a string, and an $ijk$ vertex
will correspond
to a quarter-charged monopole on an Alice string, and will play a role
in describing the topologically charged strings described below.
It is in fact possible to describe all fourth-order vertices in terms
of cubic vertices.  For instance, the magnetic monopole of figure 1 is
just the limit of the configuration shown in figure 3 as the loop
shrinks to a point.

Note that in general
a flux tube carries more information than just its holonomy.  It is
described fully by the type and direction of the flux that it carries.
The holonomy may, of course, be computed from these.
In our case, a tube carrying,
say, $j$-flux in one direction is equivalent to a tube carrying $(-j)$-flux
in the opposite direction.  It is clearly not equivalent to two $k$-tubes
and three $j$-tubes bound together, though it could have the same holonomy.

In fact, the story is a little more complicated since loops of Alice strings
can carry two types of magnetic charge.  The first of these is so-called
``Cheshire Charge''.
Before $\uem$ is broken, imagine bringing a
(magnetic) charge $q$ from infinity,
passing it through the string loop, and then returning it to infinity.  Its
charge has been changed to $-q$, and so by the conservation of magnetic charge
the Alice string must have acquired a charge $+2q$.  This is
non-localised, hence the name ``Cheshire Charge''.  Upon breaking $\pint$
to $\q$, the associated magnetic flux will be confined to tubes, and so a
Cheshire magnetically charged Alice loop will become, say, an $i$-loop with
some $k$-tubes passing through its centre.  As explained above, the direction
of the $k$-flux will reverse as we cross some gauge-dependent
branch cut.  This is illustrated in
\fig{After the superconducting phase is entered, Cheshire
magnetic charge is expressed in terms of flux tubes whose direction
reverses as they pass
through the centre of an Alice loop.}.

The second type of magnetic charge is defined as follows [\preskill]:
consider some base point $x_0$.
The holonomy of a closed path starting at $x_0$ and linking the
Alice loop is an element of $\H$.  As this path is deformed around the Alice
loop, its holonomy traces out a closed curve in $\H$.  Since $\H$ is not
simply-connected, the curve may be topologically non-trivial, and so the
Alice loop carries ``twisting'' magnetic charge.
When $\pint$ breaks to $\q$, there can no longer be smoothly varying paths in
$\H$.  Twisting magnetic charge is then manifested by quarter-monopoles
strung on the Alice string, as shown in
\fig{Twisting magnetic charge is described, in the superconducting phase,
by quarter-monopoles threaded on the Alice string.  As the path based at
$x_0$ is moved around the Alice string, its holonomy jumps each time it
crosses a $k$-tube, from $i$ to $j$ to $-i$ etc.}.
As the path based at $x_0$ is moved around the string, its holonomy jumps each
time it crosses a $k$-tube, from $i$ to $j$ to $-i$ etc.
\foot{We thank J.Preskill for discussions on this point.}
It is clear in this picture that twisting charge and Cheshire charge are
essentially equivalent --- at a finite cost in energy, the quarter-monopoles
described above could be moved together and the $k$-tubes disentangled to
leave an $i$-loop with several $k$-tubes passing through its centre.
In fact, for a non-trivial hierarchy, this will be the energetically
preferred configuration.

In this quaternionic superconducting phase, we have seen that
monopoles and Alice strings
become a network of tubes carrying $i$, $j$, and $k$ flux,
joined three-fold
\foot{As mentioned before, a four-vertex can be constructed from
three-vertices.}
at vertices, such that the total holonomy
at any vertex is trivial.  A typically ugly section of such a configuration
is shown in
\fig{Breaking $\sut \rightarrow \q$ leads to a network of $i$-, $j$-,
and $k$-strings joined at vertices such that the total holonomy at any
vertex is trivial.}.

In order to actually implement this superconducting phase,
we take the $\sut$ model of
the previous section, and add a second Higgs field $\tilde \Phi$, also
transforming in the spin-$2$ representation.
When the first Higgs field acquired an expectation value
$$\vev{\Phi}=p \diag 1 1 {-2} \,, \eqn\phivevagain$$
we saw that $\sut$ was broken to $\pint$.  If the second Higgs now gets
expectation value
$$\tilde{\vev \Phi}=\tilde p \diag 1 2 {-3} \,, \eqn\newphivev$$
then the symmetry will be further broken to $\q$.

We thus require the effective potential of $\Phi$ and $\tilde \Phi$ to
vary with temperature such that, as the Universe cools, it passes through
the following phases:

\itemitem{I)} Unbroken $\sut$ with $\vev{\Phi}=\tilde {\vev \Phi}=0$
\itemitem{II)} $\sut$ broken to $\pint$ with $\vev{\Phi}=p\  {\rm diag}
(1,1,-2)$ and $\tilde {\vev\Phi}=0$
\itemitem{III)} $\sut$ broken to $\q$  with $\vev{\Phi}= p\  {\rm diag}
(1,1,-2)$ and $\tilde {\vev \Phi}=\tilde p \  {\rm diag} (1,2,-3)$
\itemitem{IV)} Same as phase II.

At phase II, both Alice strings and monopoles are formed, and at phase III
they become the quaternionic network described above.
This network will evolve as the tension in the strings pulls the vertices
around.
In phase IV, the electromagnetic symmetry is restored, and the $k$-flux
tubes become deconfined.  The network will then dissolve into Alice
string structures and monopoles.  Notice that if we had broken $\sut$
directly to $\q$, we would have obtained the same type of network
before $\uem$ was restored.

In the original Langacker-Pi scenario, the magnetic flux in the
superconducting phase was completely confined.  This meant that to each
monopole was attached just one flux tube.  The tension in the flux tubes
then brought monopoles together very quickly, causing almost all of them
to annihilate.  In our model, the situation is more complicated.
Each monopole is attached not to one, but to four tubes, which will
generally be trying to pull it in different directions.  Thus it is not
clear whether the monopole-antimonopole pairs will be able to meet each
other sufficiently quickly to satisfy the experimental density
constraints.

In order to study this question, we must focus on the evolution of
the network formed in phase III.  The vertices will be pulled
around by the tension in the strings.  In addition, strings may
cut through each other to form new vertices;
for example, two $k$-tubes may ``escape'' from an $i$-loop to form
a new magnetic monopole as shown in
\fig{Two $k$-tubes passing through an $i$-loop can escape, forming a
magnetic monopole and an uncharged Alice loop.}.
The amplitudes for these processes will depend on the vertex masses,
and in addition must satisfy various topological constraints;
in this case it is not possible for just one $k$-tube to escape, for if
it did it would have to form a free monopole of half the elementary charge,
thus incurring Dirac's wrath.

The qualitative development of string networks of this form is likely to
depend crucially on the details of the particular model chosen.  The
effective Higgs potential will determine the string tensions and vertex
masses, which in turn control the rates of interconnection.  In addition,
the introduction of friction will change the evolution dramatically.
Not surprisingly, no thorough analysis of such networks has yet been presented.
However, numerical simulations [\numsim] have been undertaken for
simplified network models.  These indicate that, for some range of parameters,
a network's mass will not come to dominate the Universe.  For our
applications we require far more than this --- we need the network to
have become sufficiently dilute that after restoration of $\uem$ the
surviving magnetic monopoles satisfy the experimental density bounds.
Thus the question of whether the model so far described can solve the
monopole problem will need a more thorough knowledge of network evolution.
However, in section 5 we shall present a
symmetry breaking scenario in which it is clear that the monopole
density can easily be reduced to the desired level.  It is this that
we regard as definitively solving the monopole problem in an Alice model.

\hfill

\chapter{Breaking to Smaller Groups}

In the previous section we considered a superconducting phase in which $\pint$
was broken to $\q =\{ \pm 1, \pm i, \pm j, \pm k \}$.
As a natural extension of this, we
ask what happens if instead we break to
some subgroup of $\q$.
Up to conjugacy, there are three such subgroups:
$\zf=\{ \pm 1, \pm i \}$, $\zt=\{ \pm 1 \}$, and the trivial group.

For ease of visualisation, suppose that we first break $\pint$ to $\q$, and
then break this further in stages.
By adding a Higgs field ${\bf v}$ in the spin-$1$ representation, we can
break the symmetry to $\zf$.  After $\tilde {\vev{ \Phi}}$ has become
$$\tilde {\vev \Phi}=\tilde p \diag 1 2 {-3} \,, \eqn\phivevr$$
we let ${\bf v}$ acquire the expectation value
$$\vev{{\bf v}}=\vector v 0 0 \,. \eqn\vvev$$
This breaks $\q$ to $\zf=\{ \pm 1, \pm i \}$.
At this point, strings of holonomy $\pm j$, $\pm k$ will no longer be allowed,
and so they must pair up.
\foot{The pairing-up mechanism can be understood in the following way.
As the symmetry breaks, the $j$- and $k$-strings will become boundaries
of domain walls on which $\vev{{\bf v}}$ remains zero.
The tension in these walls, aided by string interconnections, causes them
to collapse.  As they do so they will bring
pairs of $j$- and $k$-strings together.  This will occur on a time-scale that
is short compared to the subsequent evolution of the string network.}
A magnetic monopole will now have two $(-1)$-holonomy
flux tubes attached to it.
Note that, as before, a $1$-string will decompose into smaller flux
fractions.  Hence the network will consist of $i$-strings
and $(-1)$-strings, with $i^2 (-1)$ vertices,
and beads strung on $(-1)$ strings.
The beads will interpolate between the various kinds of flux that
the $(-1)$-strings can carry.
Like our previous quaternionic network, this one will evolve by the tension in
the strings accelerating the vertices, and by strings interconnecting with
each other to form new vertices.  Again, the evolution will depend
on the details of the model, and the question of whether experimental
monopole bounds can be satisfied must await more refined network analyses.

If we add a further Higgs field ${\bf v}'$ in the spin-$1$ representation,
and let it acquire expectation value
$$\vev{{\bf v}'}=\vector 0 {v'} 0 \,, \eqn\vprimevev$$
then we see that $\zf$ is further broken to $\zt=\{ \pm 1 \}$.
Since strings of holonomy $i$ are not allowed, they too must pair up.
In contrast to the previous models, holonomy-$1$ strings can now be stable,
since an $ijk$-string is unable to break up into anything smaller.
Hence the network will consist of $(-1)$- and $1$-strings carrying
various types of flux, and with beads strung on them.  There will be
$1^3$ and $(-1)^21$ vertices, and $1$-tubes will be allowed to end (at
generalised monopoles). No matter how low the hierarchy, a string of
non-trivial holonomy can never break.  However, a $1$-string can break,
forming a generalised monopole-antimonopole pair.  Thus, in this example,
if the $\pint \rightarrow \zt$ phase transition occurs soon after the
original $\sut \rightarrow \pint$ transition, then the $1$-strings will
dissolve, leaving just a $(-1)$-string network.
Magnetic monopoles, though, have two
$(-1)$-strings attached to them, and so as before it is not clear from
existing analyses that the subsequent network evolution
will be fast enough to get the
monopole density down to experimentally allowed levels.

In order to break $SU(2)$ completely, we add a doublet
Higgs field, $\psi$.
This is allowed to acquire the expectation value
$$\vev{\psi}=\left( \matrix{a \cr 0 \cr } \right) \,, \eqn\psivev$$
which is obviously not left invariant by any non-trivial subgroup of $\sut$.
Now strings of holonomy $(-1)$ are not allowed, and so they must pair up.
This will leave just $1$-tubes, carrying various types of flux, strung
with beads, and joined together at vertices.  The tubes will be
able to end, either on magnetic monopoles or on other generalised
monopoles whose
nature will depend on the flux that they carry.

If the hierarchy is sufficiently low, then the entire network will dissolve
almost immediately, leaving behind no topological structures.  This case is
as destructive to Alice strings as inflation, and hence is of no
interest to us.  We will assume, then, that the network does not polarise
too quickly.  In this case it will still exhibit a major difference from the
partial symmetry breaking networks so far described; each monopole will
now be connected to exactly one flux tube.  This was the essential property
of the usual Langacker-Pi mechanism, and so it may seem that the monopole
problem would be solved.  However, there will also be loops of string that
are multiply magnetically charged.  These will have many flux tubes attaching
them to the rest of the network, and so they will move only slowly.  At the
termination of the superconducting era, we will be left with magnetically
charged loops of Alice string which may subsequently
break up and contract away to points, giving birth once again to the
unwanted magnetic monopoles.  The actual magnitudes of these effects are
unknown and await a more detailed analysis.

We have seen that breaking the symmetry to smaller subgroups leads to
the same sort of network that we found in the last section.
When $1$-strings are topologically stable, some qualitatively new
features emerge, namely the breaking of strings and the fast motion
of monopoles.
Though it may possibly be found from more complete analyses that some
of these networks {\it can} solve the monopole problem, we do not
claim to have reached any such conclusion here.  However, in the
next section we shall present a new model containing Alice strings.
Even without detailed knowledge of its evolution, it can clearly
be seen to eliminate magnetic monopoles in the desired way.

\hfill

\chapter{Annihilating Monopoles Faster than Alice Strings}

With the deliberations of the previous section in mind, we shall now
develop a model in which monopoles are quickly annihilated to leave
an Alice string network.  We have seen that the only flux tubes that
disappear quickly are those with trivial holonomy; if the hierarchy
is low, they will dissolve into monopole-antimonopole pairs.
Thus it would seem desirable that the magnetic flux be completely
confined into $1$-tubes soon after the monopoles are formed.
However, if we {\it completely} break the symmetry then the Alice strings
will also be dissolved.  To avoid this
we need magnetic flux to be completely confined,
and Alice flux to be only partially confined.
This could never have occured for the $\sut \rightarrow \pint$ model,
since any non-trivial subgroup of $\q$ contains the element $(-1)$, and so
any symmetry breaking that does not confine all flux would leave
at least two flux tubes attached to each monopole.

To achieve our goal, then, we shall have to construct a new model
containing Alice strings.  For this we
consider an $\suthree$ gauge theory, with a Higgs field $\Lambda$
in the ${\bf 6}=(2,0)$ representation.  We can regard $\Lambda$ as being
a symmetric $3 \times 3$ matrix with the transformation law
$$\Lambda \rightarrow M \Lambda M^T \qquad M \in \suthree \,.
\eqn\Lambdatrans$$
Now let $\Lambda$ acquire the unitary gauge expectation value
$$\vev{\Lambda}=h \diag 2 2 1  \,. \eqn\Lambdavev$$
It is easy to see that this breaks $\suthree$ down to $\ot$, and so
we have a model containing Alice strings.  Note that since $\suthree$
is simply-connected, we don't need to consider its covering group.
This is the crucial difference from the $\sothree \rightarrow \ot$
model.

As before, the symmetry breaking will produce monopoles, since
$$\Pi_2(\G / \H )=\Pi_2(\suthree/ \ot )=\Pi_1(\ot)\ne0  \,. \eqn\pies$$
In order to enter the superconducting phase, $\ot$ must be broken down
to some discrete subgroup.  As before, for ease of visualisation,
we shall imagine doing this in stages.  First we break the symmetry
to the dihedral group $\dt=\{ 1,a, b,c \}$, consisting of rotations
by $\pi$ about each of the principal axes of $\Lambda$.  This may be achieved
by introducing a second Higgs field $\tilde \Lambda$, also in the $\bf 6$
representation, and letting it acquire expectation value
$$\tilde{\vev \Lambda}=\tilde h \diag 3 2 1 \,, \eqn\newchivev$$
in the unitary gauge.  At this point, Alice strings have become $a$- and $b$-
flux tubes, and magnetic flux is confined into $c$-flux tubes.  Each
monopole will have two $c$-tubes attached to it.

Now we break $\dt$ down to $\zt=\{ 1,a \}$.  This can be done
by introducing a Higgs
field ${\bf v}$ in the ${\bf 3}=(1,0)$ representation of $\suthree$ and letting
it acquire the (unitary gauge) expectation value
$$\vev{{\bf v}}=\vector v 0 0 \,. \eqn\thisvvev$$
No longer will $b$- and $c$- flux tubes be allowed to exist in isolation,
and so they will pair up.  In particular, a magnetic monopole will
now have a single $1$-tube tied to it, carrying flux $c^2$.
We will thus have a network of strings of holonomy $1$ and $a$, as shown
in
\fig{Breaking $\suthree \rightarrow \zt=\{ 1,a \}$ leads to a network
of $1$-strings and $a$-strings, with each magnetic monopole attached to
the end of a $1$-string.}.
Now suppose that the hierarchy is low.  Then the $1$-strings will dissolve
into monopole-antimonopole pairs which immediately annihilate.
This will leave only strings of holonomy $a$.
There are two types of these --- ones carrying a unit of $a$-flux, and
ones carrying both $b$- and $c$- flux.  The latter simply consist of a
$b$-string and a $c$-string bound together.
Because of the symmetry between $a$, $b$ and $c$, it is clear that
$a$-strings have approximately half the tension of $bc$-strings.
Thus, since the
hierarchy is low, the latter will decay into the former, via the process
shown in
\fig{A $bc$-string has approximately twice the tension of
an $a$-string, and so,
if the hierarchy is low, sections of it will be replaced by sections
of $a$-string.  These will then expand until the entire $bc$-string
has been transformed into an $a$-string.}.
This will leave just $a$-strings sewn with beads at which the direction
of the $a$-flux changes.  There will be no stable vertices joining more
than two $a$-strings.
At the end of the superconducting era, then, we see that a relative
abundance of Alice strings will remain.
Moreover, these strings will be devoid of magnetic charge,
and so there
is no danger of them later decaying to give re-birth to monopoles.

Kibble and Weinberg [\kibbwein] have pointed out a variant of the
Langacker-Pi mechanism in which, rather than forming monopoles and
then annihilating them, their initial production is prevented.
\foot{This argument is valid when the symmetry breaking does not
give rise to a network of stable cosmic strings with non-trivial
vertices.  In this case topological structures may be surrounded
by spheres on which there are no singularities.}
We can implement this scenario by breaking the symmetry directly
to $\zt$ in the following way.
Let $\Lambda$ and ${\bf v}$, as before, lie in the
$\bf 6$ and $\bf 3$ representations
of $\suthree$, and let the Higgs pair $(\Lambda,{\bf v})$ acquire the
expectation value
$$\vev{(\Lambda,{\bf v})}=\left( \diag {2h} {2h} h , \vector v 0 0 \right)
\eqn\Lambdavvev$$
in the unitary gauge.
This breaks $\suthree$ down to $\zt$.  In a non-singular gauge, then, we
have
$$\vev{(\Lambda,{\bf v})}=
\left( M \diag {2h} {2h} h M^T, M \vector v 0 0 \right)
\,, \qquad M \in \suthree / \zt \,. \eqn\nonsing$$
Since $\Pi_2(\suthree / \zt)=\Pi_1(\zt)=0$, we can see that $M$,
viewed as a $3\times 3$ matrix,
is topologically trivial on a sphere ${\rm S}^2$, though not on a
circle.

Next we let $\vev{{\bf v}} \rightarrow 0$,
while leaving $\vev {\Lambda}$ unaltered.
Then we have
$$\vev{\Lambda}=M \diag {2h} {2h} h M^T \,, \qquad M\in \suthree / \ot \,.
\eqn\Lambdaokvev$$
but still $M$ is topologically trivial on ${\rm S}^2$, and so there
are no monopoles.  Summarising, the phases of symmetry breaking that we have
described are:
\itemitem{I)} $\suthree$ unbroken.
\itemitem{II)} $\suthree$ broken to $\zt$.
$\vev{(\Lambda,{\bf v})}=\left( \diag {2h} {2h} h , \vector v 0 0 \right)$
\itemitem{III)} $\suthree$ broken to $\ot$.  $\vev \Lambda \ne 0$,
$\vev {\bf v}=0$.

And in phase III we are left with Alice strings but no monopoles.

This mechanism for preventing the formation of monopoles is really just the
limit of the previous model as the hierarchy is taken to zero.  In any case,
either of the scenarios of this section can easily account for an abundance
of Alice strings with a paucity of monopoles.

\hfill

\chapter{Conclusion}

In this paper we have studied the Langacker-Pi mechanism as a means of
solving the monopole problem in models containing Alice strings.
We found that the superconducting phase transition gives rise to a
network of flux tubes which evolves in a complex way.  It is difficult to
analyse this evolution in detail, and we have not done so here.  However,
for the particular symmetry breaking pattern $\uz \rightarrow \zt$ with
a low hierarchy, we have shown that magnetic monopoles are quickly
annihilated, leaving behind a network of Alice strings.

This observation opens up the possibility that Alice strings were formed
after any inflationary eras, and hence could have important cosmological
effects.  Since our Universe today has a vacuum which is not symmetric
under C,
we must postulate that this is the result of some spontaneous symmetry
breaking of a GUT gauge group containing C.  When this occured, each
Alice string became the boundary of a
domain wall.  These domain walls are possibly superconducting, and have
the property that they can act as ``filters'' that convert antimatter
into matter.  This could account for the baryon
asymmetry of the Universe.  We shall investigate these matters further in
a separate publication [\dogpiss].

\ack{
We would like to thank J.Preskill for many invaluable discussions.}
\refout
\figout
\bye